\documentclass{elsart}    
\usepackage{graphicx}
\usepackage{amssymb}
\begin{document}

\begin{frontmatter}

\title{Band structure related wave function symmetry of amphoteric Si dopants in GaAs}
\author[goe]{S. Loth}
\author[goe]{M. Wenderoth\corauthref{cor1}}
\author[goe]{K. Teichmann}
\author[goe]{R. G. Ulbrich}
\address[goe]{IV. Physikalisches Institut der Georg-August-Universit\"at G\"ottingen, Friedrich-Hund-Platz 1, 37077 G\"ottingen, Germany}
\corauth[cor1]{Corresponding author: wendero@ph4.physik.uni-goettingen.de}

\begin{abstract}
Autocompensated Si-doped GaAs is studied with cross-sectional scanning tunneling spectroscopy (X-STS). The local electronic contrasts of substitutional Si(Ga) donors and Si(As) acceptors under the (110) cleavage plane are imaged with high resolution. Si(Ga) donor atoms exhibit radially symmetric contrasts. Si(As) acceptors have anisotropic features. The anisotropic acceptor contrasts are traced back to a tunnel process at the valence band edge. They reflect the probability density distribution of the localized acceptor hole state.
\end{abstract}

\end{frontmatter}

\section{Introduction}
The local electronic properties of dopant atoms in semiconductors became accessible with the advent of cross-sectional scanning tunneling microscopy (X--STM).
In such experiments a bulk semiconductor sample is cleaved along one of the crystal's preferred cleavage planes and an atomically flat cross-section through the sample is laid open, which allows the tip to probe not only surface atoms but also buried dopants.
In III-V semiconductors a variety of different donor and acceptor species has been investigated.
For donors only radially symmetric features have been reported \cite{zhe94si,dep01,fee02,kor01sn}.
All investigated acceptors exhibit distinct anisotropic features for certain tunneling conditions\cite{zhe94,kor01,ars2003,yak04,mah05,lot06j,mar2007b}. Either triangular shaped or bow-tie like contrasts are imaged. These acceptor contrasts are linked to the host lattice symmetry and its pronounced non-spherical valence band structure\cite{lot06j,mar2007b}. They are mirror-symmetric over the \{110\} mirror planes and have an asymmetry with reference to (001).
Earlier it was suggested that whether anisotropic contrasts are observed or not depends on chemical characteristics of each doping element, e.g., electronic orbitals\cite{zhe94} or tetragonal strain fields surrounding the dopant\cite{kor01}. Recent studies indicate that the anisotropic shapes relate to intrinsic properties of the semiconductor matrix rather than a property of the specific doping element\cite{mar2007b,mon2006,lot06p}.

This work focusses on a comparative spectroscopic study of acceptor and donor related contrasts for the same doping element. The amphoteric dopant silicon in GaAs\cite{sch93} is an ideal candidate for this experiment. Si can be substituted either on the Ga site, where it becomes a shallow donor, or on the As site, where it is a shallow acceptor. Although Si is predominantly incorporated as donor, the significant formation of Si(As) acceptors occurs for doping levels exceeding $\sim2\times10^{18}cm^{-3}$ \cite{cha1981,nea1983,dom96}. In such autocompensated samples the doping consists of donors and acceptors that are intermixed in the same volume of the crystal. In one scanning tunneling spectroscopy (STS) measurement donors and acceptors can be compared directly.

\section{Experiment}
\begin{figure}
\includegraphics[scale=0.73]{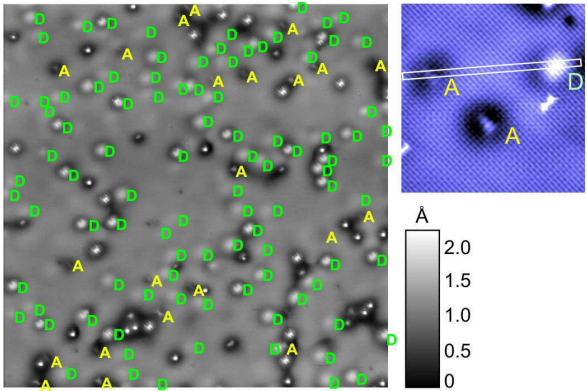}
\caption{\label{dopident} (Color online)
(100 $\times$ 100)~nm$^2$ constant current topography of the autocompensated GaAs sample. Sample bias -1.8 V, tunnel current 50pA. Donors are marked with green (D)s and acceptors with yellow (A)s. small image: zoom in on a region with two acceptors and one donor. The section in Fig.~\ref{ad} is evaluated along the line indicated by the white rectangle.}
\end{figure}
The experiments are performed in a low temperature STM operating in UHV at a base pressure better than $2\times10^{-11}$ mbar. Details of the experimental setup are given in ref.\cite{lot07prb}. The GaAs samples are cleaved \textit{in situ} at room temperature and they are transferred to the precooled microscope where they reach the equilibrium temperature of 8~K within less than an hour after cleavage. The sample's nominal doping concentration of $6.5\times10^{18}cm^{-3}$ silicon is validated by large scale STM topographies. Figure~\ref{dopident} presents a (100 $\times$ 100)~nm$^2$ constant current topography acquired on a (110) cleavage plane at $-1.7$~V sample bias. At this bias, subsurface dopants show prominent topographic contrasts. Donors and acceptors are easily identified by their charge signature (see Fig.~\ref{ad})\cite{dom98}. The dopant distribution in the STM-image is nearly "homogenous" and the different doping species are intermixed which is an important prerequisite for the following comparative studies.
A total of 94 donors and 22 acceptors in five different layers are identified. Within the accuracy of this analysis the concentration of electrically active donors is $4.4 - 6.8 \times 10^{18}cm^{-3}$ and the acceptor concentration is in the range of $1.0 - 2.3 \times 10^{18}cm^{-3}$. This demonstrates that significant autocompensation via the formation of acceptors was achieved in the presented sample. Nevertheless, the sample is still highly n-conducting which was confirmed by macroscopic van der Pauw and Hall measurements. The free n-type carrier concentration is $~3.6\times 10^{18}cm^{-3}$, so the Fermi energy $E_F$ is still well within the conduction band.

The positions of the dopant atoms under the surface are marked by white circles in Fig.~\ref{ad} and Fig.~\ref{didu}. They are determined by the center-of mass of the radially symmetric contrasts in STM topographies at high positive and large negative sample biases\cite{zhe94si,dep01,kor01}.

The following sections focus on spectroscopic measurements of acceptors and donors that are below the (110) surface. I(V)-characteristics are recorded at every scan position of the small (blue-white colored) topography in Fig.~\ref{dopident}. The resulting three dimensional tunnel current data set $I(x,y,V)$ is numerically differentiated (details of the procedure are given in ref.\cite{lot07prb}). A section through this dataset cut along a line $s$ on the topography gives information of the energy dependent distribution of localized and extended states in the sample. In such sections the differential conductivity $dI/dV$ is normalized with a smoothed total conductance $(\overline{I/V})$. This yields a measure of the sample's local density of states [LDOS(V)] \cite{fee87,wie1994}. Such a  $[dI/dV/(\overline{I/V})](s,V)$ section is shown in Fig.~\ref{ad}c.
Maps of the $[dI/dV](x,y)$ signal for constant sample bias contain information of the spatial distribution of certain conductivity peaks that have been identified, e.g., in the $dI/dV/(\overline{I/V})$ sections. In these maps the $I(x,y,V)$ dataset is normalized to a plane of constant height ($z=z_0=const.$) rather than to total conductance to eliminate crosstalk from the topographic signal\cite{gar04}. Figure~\ref{didu} presents the resulting $dI/dV|_{z_0}$ maps.

The tip-sample interaction has to be considered to derive information from STS measurements : the tip induces bias dependent space-charge layers at the surface that locally shift the sample's bands which results in the tip-induced band bending [TIBB(V)] at the surface\cite{fee87}. Knowledge of the precise bias value for which the band bending vanishes is important. Higher bias results in a depletion layer at the surface, i.e., upward bending of the bands. Lower sample bias results in an accumulation layer in the conduction band. Based on the experimental determination of the tip work function\cite{lot07prb} (4.3~eV in the presented STS-measurement) the flat band bias is fixed to $-0.2$~V bias. The complete TIBB(V) dependence is then numerically evaluated\cite{fee87,raa02,fee03}.

\section{Results}
\begin{figure}
\includegraphics[scale=0.87]{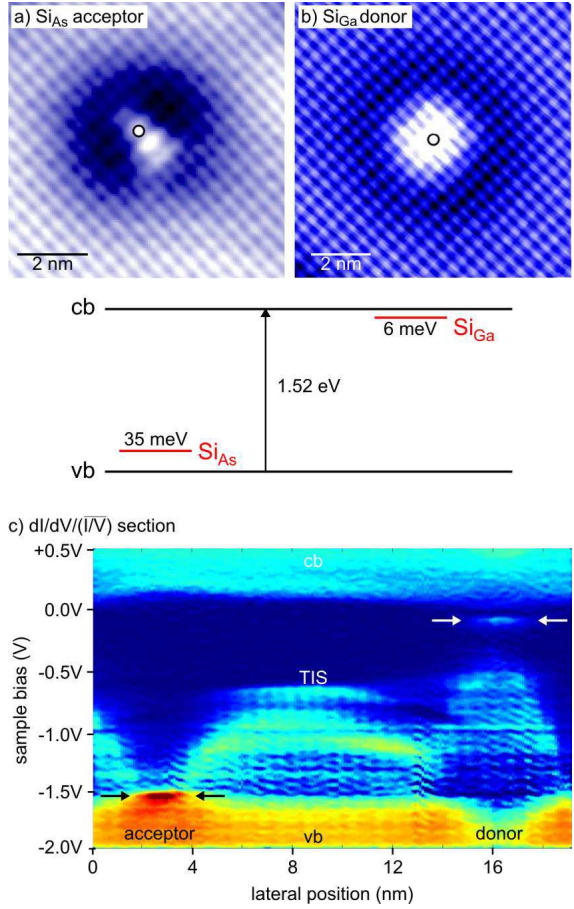}
\caption{\label{ad} (Color online)
upper part: Constant current topographies of (a) Si(As) acceptor and (b) Si(Ga) donor both acquired at -1.8~V sample bias and 100~pA tunnel current. The projected positions of the dopant atoms under the surface are marked by white circles. middle part: band diagram sketch of the binding energies of both dopant species relative to conduction band (cb) and valence band (vb) edge. lower part: (c) $dI/dV/(\overline{I/V})$ section through an acceptor and a donor (see Fig.~\ref{dopident}) plotted color coded against sample bias and lateral position (low conductivity is blue and high conductivity is cyan - red).}
\end{figure}
The dopant induced features in constant current topographies are analyzed in Fig~\ref{ad}. It shows filled states images of a Si(As) acceptor and a Si(Ga) donor. Both topographies are acquired at  $-1.8$~V. The conduction band is pulled under the Fermi energy. Electrons are accumulated at the surface and the C3 surface resonance is detected above the undisturbed surface (seen as lines running along the [001] direction)\cite{che79,raa02}.
The donor (Fig.~\ref{ad}b) appears as a circular protrusion with an extension of about 4~nm. It is centered on the projected dopant atom position. The protrusion is surrounded by a dark halo which is the topographic image of charge density oscillations (CDOs) around the donor in the nearly free electron gas of the accumulation layer\cite{wie96,wen99,dom99}. The donor is screened by the tip induced states (TIS) of the accumulation layer. These observations are in good agreement with previous reports of homogenously n-doped samples\cite{zhe94si,dep01,fee02}.
In contrast to that, the acceptor shows up at $-1.8$~V as a circular depression superimposed with an elongated nearly triangular protrusion. The circular depression persists for the whole investigated bias voltage range from $-2.5$~V to $+3.0$~V. It is comparable in size with the protrusion above the donor and also centered on the dopant atom. The triangular feature is shifted to one side of the dopant's position. It only appears in a narrow energy interval around  $-1.6$~V to $-1.9$~V. This may be the reason why the triangular contrast above the Si(As) acceptor has not been reported in literature up to now, while the circular depression was\cite{dom98,dom96}.
As a first result, these measurements demonstrate that the amphoteric dopant can exhibit either circular symmetric or anisotropic contrasts depending on the dopant's configuration as a donor or acceptor in the semiconductor matrix.

The next step is to identify the energetic origin of the anisotropic feature above the acceptors. The band diagram in Fig.~\ref{ad} sketches the bulk energetic conditions for both dopants. The Si(Ga) donor has a nominal bulk binding energy of 6~meV and is below the conduction band edge (cb)\cite{Liu1988}. The Si(As) acceptor is close to the valence band edge (vb) and has a binding energy of 35~meV\cite{Ash1975}.
A $dI/dV/(\overline{I/V})$ section gives insight in the energetic conductivity distribution of the contrasts. Figure~\ref{ad}c presents a cut directly through one acceptor and donor contrast as seen in the small topography in Fig.~\ref{dopident}. The onset of conduction band tunneling is detected directly above $0.0$~V sample bias and valence band tunneling at $-1.55$~V. The measured band tunneling onsets closely match with the above sketched bulk values. This is due to the sample's high doping level. The tip-induced band bending (TIBB) is small and the band tunneling is easily observed\cite{fee03}.

The donor exhibits a conductivity peak at $-0.10$~V. Above the acceptor a prominent conductivity peak is detected at $-1.54$~V. The onset of tip induced states (TIS) in the accumulation layer is observed at $-0.6$~V \cite{wen99,dom99} between the dopant atoms. The TIS persist down to large negative voltage but they are masked by the valence band tunneling for bias below $-1.54$~V. While the TIS are also present in the region of the donor, they are not visible above the acceptor atom.

\begin{figure*}
\includegraphics[scale=0.81]{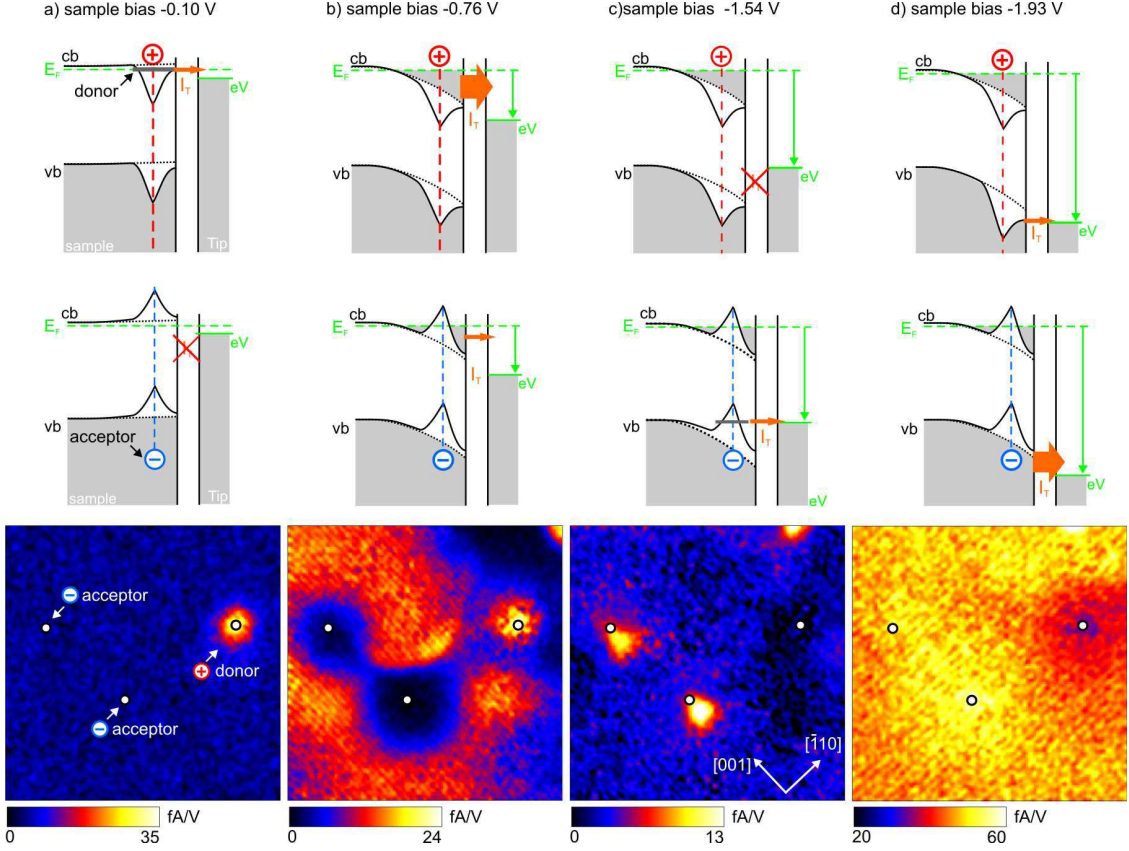}
\caption{\label{didu} (Color online)
Rigid-band diagrams of four characteristic bias voltages in the STS measurement of Fig.~\ref{ad}c. The labels cb, vb, E$_F$, eV and I$_T$ are described in the main text. The sample bias dependent electronic configuration at the donor is shown in the upper row and for the acceptor it is sketched in the middle row. The orange arrows (I$_T$) indicate the tunnel current contributing to the orange colored differential conductivity in the respective dI/dV maps (lower row). The chosen (20~$\times$~20)~nm$^2$ region for the $dI/dV|_{z_0}$ maps is the same as the one in Fig.~\ref{dopident} (small image) and shows two subsurface Si(As) acceptors and one Si(Ga) donor. The four characteristic voltages are: (a) $-0.10$~V, imaging of the donor state's probability distribution. (b) $-0.76$~V, delocalized conductivity of the tip-induced states (TIS). (c) $-1.54$~V, imaging of the acceptor state. (d) $-1.93$~V, valence band tunneling.}
\end{figure*}

The close matching of $dI/dV/(\overline{I/V})$ section and the band diagram gives rise to the assumption that the conductivity peak at $-1.54$~V at the acceptor is an image of the acceptor state and the conductivity peak at $-0.1$~V above the donor relates to the donor state.
This is elucidated in detailed analysis. Fig.~\ref{didu} presents conductivity maps (dI/dV-maps) for four characteristic bias voltages of the same sample region as shown in Fig.~\ref{dopident} (small topography).
Two Si(As) acceptors and one Si(Ga) donor are visible in the imaged area.
Band diagrams for each dI/dV-map depict the respective electronic configuration above donor and acceptor. The band alignment is sketched for each of the four situations according to the calculated TIBB(V) dependence. The conduction band onset and valence band onset are labeled (cb) and (vb) respectively. The Fermi energy (E$_F$) is drawn as dotted green line. The sample bias (eV) shifts the tip and sample Fermi energy with respect to each other which is indicated as solid green line at the tip. Filled states are drawn in grey, and empty states are white. The tip induced band bending (TIBB) creates the parabolic slope of the band edges from bulk to surface. The upper scheme of each column in Fig.~\ref{didu} shows the configuration for the donor and the lower one is sketched for the acceptor atoms. The major difference between them is the local charge of the dopants as indicated by the Coulomb potential like distortion of (cb) and (vb). The acceptor possesses a negative core charge (-) whereas the donor core is positively charged (+). The tunneling process leading to the contrast above the respective dopant is emphasized by orange colored tunnel current arrows (I$_T$). The thickness of each arrow indicates the strength of the respective tunnel process.

The first characteristic bias voltage is $-0.10$~V. According to the $dI/dV/(\overline{I/V})$ section in Fig.~\ref{ad}c this is the spatial distribution of the localized state above the donor. The dI/dV-map in Fig.~\ref{didu}a shows a circular region of enhanced conductivity above the donor atom. The rest of the sample shows no conductivity. The band bending situation (rigid-band model Fig.~\ref{didu}a) identifies the observed contrast: at $-0.10$~V the sample is still in depletion, i.e., the bands are slightly bent upward.
In this configuration the donor level crosses the Fermi energy, an electron is bound to the positive dopant core and contributes to the tunnel current. Thus, the circular shaped conductivity directly reflects the probability density distribution of the donor wave function. It is imaged as a circular symmetric contrast with about 5~nm diameter.

Fig.~\ref{didu}b presents the dI/dV-map acquired at $-0.76$~V. Delocalized conductivity is detected on the whole sample. For such voltages the sample is in accumulation. The TIBB is negative, the conduction band is pulled under the Fermi energy and electrons are accumulated. The atomic corrugation of the delocalized conductivity resembles the C3 surface resonance in the conduction band\cite{raa02}. This indicates that the imaged conductivity is at the conduction band. The above mentioned tip induced states (TIS) are visualized. The dI/dV-map shows charge density oscillations of the nearly free electron gas\cite{wie96,wen99,dom99}. Both donor and acceptor represent scattering centers within this electron gas. As depicted in the rigid-band diagrams the donor exhibits an attractive potential and the conductivity is locally enhanced. In contrast to that the negative charge at the acceptor atom is a repulsive scattering center and locally repels the electrons. The delocalized conductivity seen as orange color on the undisturbed surface does not reach the acceptor atom. A depression of about 4 nm width around the acceptors is visible.

Figure~\ref{didu}c shows the dI/dV-map acquired at $-1.54$~V, the bias voltage of the pronounced localized state above the acceptor (compare with Fig.~\ref{ad}c). The conductivity on the undisturbed surface and above the donor is low (black and blue color). Enhanced conductivity is detected only at the acceptors. The image demonstrates that the corresponding state has a pronounced anisotropic triangular shape. Additionally, the conductivity is shifted to the [00$\overline{1}$] side of the dopant atom. Above the acceptor in the upper left corner a faint branch of conductivity is detected on the [001] side as well.
No valence band states are accessible directly at the surface due to the negative band bending. According to the band diagrams the anisotropic conductivity is an image of the acceptor state that is resonantly aligned with the valence band edge. In analogy to the image of the donor wave function at $-0.10$~V (Fig.~\ref{didu}a) the anisotropic contrast reflects the probability density distribution of the acceptor hole state.
It extends about 5~nm along [001] and 3~nm along [$\overline{1}$10] and highlights the pronounced asymmetry of the valence bands at the sample surface.

When high negative bias exceeding $-1.9$~V is applied (Fig.~\ref{didu}d), the sample bias overcomes the band gap plus the TIBB. Tunneling out of the valence band states directly at the surface is possible. Conductivity spreads over the whole sample area. It has the atomic corrugation of the A5 surface resonance of the valence band (seen as lines running through the image along [$\overline{1}$10])\cite{che79,Ebe1996}. The dopants induce broad changes in the local conductivity due to their charge. As depicted in the rigid-band diagrams the negative charge of the acceptor core shifts the bands upward. More states are available for tunneling at the surface and increased conductivity is detected. The positive charge of the donor core pulls down the bands and the conductivity is decreased.

The dI/dV-maps allow decomposing the topographic contrasts for donor and acceptor.
Above the Si(Ga) donor the previously reported features, i.e., circular symmetric donor state and circular symmetric CDOs, could be reproduced\cite{fee02}. It is worth noting that the variety of studies on Si(Ga) donors leaves but little doubt that the donor exhibits no anisotropic contrasts.
The images of Si(As) acceptors on the other side consist of the superposition of two tunnel processes at different bands: The negative core charge repels the electrons of the TIS in the conduction band which results in a circular depression in the topography. The nearly triangular protrusion within the circular depression originates from the additional tunnel process at the valence band edge. It is only visible when the acceptor state is aligned with the valence band and the sample bias exceeds the GaAs band gap energy.  The anisotropic shape is attributed to the anisotropic probability density distribution of the acceptor hole state.

\section{Conclusion}
In summary we used spatially resolved I(V)-spectroscopy to comparatively study the spectroscopic features of donors and acceptors in GaAs. The choice of an autocompensated sample in which acceptors and donors are formed by the same doping element rules out the possibility that anisotropic shapes are related to specific properties of each doping element. The observation of anisotropic contrasts has one basic prerequisite: the presence of a localized state close to the valence band edge.

\section{Acknowledgments}
We gratefully acknowledge U. Kretzer, Freiberger Compound Materials GmbH, for providing the samples.
This work was supported by DFG-SFB 602 A7, DFG-SPP 1285 and the German National Academic Foundation.




\end{document}